\documentclass[]{aa}

\usepackage{amsmath}
\usepackage[dvips]{graphicx}
\usepackage{natbib}
\bibpunct{(}{)}{;}{a}{}{,}
\usepackage[american]{babel}

\def\ga{\mathrel{\mathpalette\fun >}}
\def\fun#1#2{\lower3.6pt\vbox{\baselineskip0pt\lineskip.9pt
  \ialign{$\mathsurround=0pt#1\hfil##\hfil$\crcr#2\crcr\sim\crcr}}}

{\catcode`\|=\active
  \gdef\Braket#1{\left<\mathcode`\|"8000\let|\bravert {#1}\right>}}
\def\bravert{\egroup\,\vrule\,\bgroup}

\def\gsim{\lower.5ex\hbox{\gtsima}}
\newfont{\mc}{cmcsc10 scaled\magstep2}
\newfont{\cmc}{cmcsc10 scaled\magstep1}

\newcommand{\bgi}{\begin{itemize}}
\newcommand{\edi}{\end{itemize}}

\newcommand{\be}{\begin{equation}}
\newcommand{\ee}{\end{equation}}

\newcommand{\bea}{\begin{eqnarray}}                  %
\newcommand{\eea}{\end{eqnarray}}                    %

\newcommand{\beaa}{\begin{eqnarray*}}                %
\newcommand{\eeaa}{\end{eqnarray*}}                  %

\newcommand{\bgd}{\begin{description}}
\newcommand{\edd}{\end{description}}

\newcommand{\bgf}{\begin{figure}}
\newcommand{\edf}{\end{figure}}

\newcommand{\bgc}{\begin{center}}
\newcommand{\edc}{\end{center}}

\newcommand{\bgt}{\begin{tabular}}
\newcommand{\edt}{\end{tabular}}

\newcommand{\bge}{\begin{enumerate}}
\newcommand{\ede}{\end{enumerate}}

\DeclareRobustCommand{\ion}[2]{%
 \relax\ifmmode
 \ifx\testbx\f@series
 {\mathbf{#1\,\mathsc{#2}}}\else
 {\mathrm{#1\,\mathsc{#2}}}\fi
 \else\textup{#1\,{\mdseries\textsc{#2}}}%
 \fi}

\def\jnl@style{\it}
\def\aaref@jnl#1{{\jnl@style#1}}

\def\aaref@jnl#1{{\jnl@style#1}}

\def\aj{\aaref@jnl{AJ}}                   
\def\araa{\aaref@jnl{ARA\&A}}             
\def\apj{\aaref@jnl{ApJ}}                 
\def\apjl{\aaref@jnl{ApJ}}                
\def\apjs{\aaref@jnl{ApJS}}               
\def\ao{\aaref@jnl{Appl.~Opt.}}           
\def\apss{\aaref@jnl{Ap\&SS}}             
\def\aap{\aaref@jnl{A\&A}}                
\def\aapr{\aaref@jnl{A\&A~Rev.}}          
\def\aaps{\aaref@jnl{A\&AS}}              
\def\azh{\aaref@jnl{AZh}}                 
\def\baas{\aaref@jnl{BAAS}}               
\def\jrasc{\aaref@jnl{JRASC}}             
\def\memras{\aaref@jnl{MmRAS}}            
\def\mnras{\aaref@jnl{MNRAS}}             
\def\pra{\aaref@jnl{Phys.~Rev.~A}}        
\def\prb{\aaref@jnl{Phys.~Rev.~B}}        
\def\prc{\aaref@jnl{Phys.~Rev.~C}}        
\def\prd{\aaref@jnl{Phys.~Rev.~D}}        
\def\pre{\aaref@jnl{Phys.~Rev.~E}}        
\def\prl{\aaref@jnl{Phys.~Rev.~Lett.}}    
\def\pasp{\aaref@jnl{PASP}}               
\def\pasj{\aaref@jnl{PASJ}}               
\def\qjras{\aaref@jnl{QJRAS}}             
\def\skytel{\aaref@jnl{S\&T}}             
\def\solphys{\aaref@jnl{Sol.~Phys.}}      
\def\sovast{\aaref@jnl{Soviet~Ast.}}      
\def\ssr{\aaref@jnl{Space~Sci.~Rev.}}     
\def\zap{\aaref@jnl{ZAp}}                 
\def\nat{\aaref@jnl{Nature}}              
\def\iaucirc{\aaref@jnl{IAU~Circ.}}       
\def\aplett{\aaref@jnl{Astrophys.~Lett.}} 
\def\apspr{\aaref@jnl{Astrophys.~Space~Phys.~Res.}}
\def\bain{\aaref@jnl{Bull.~Astron.~Inst.~Netherlands}} 
\def\fcp{\aaref@jnl{Fund.~Cosmic~Phys.}}  
\def\gca{\aaref@jnl{Geochim.~Cosmochim.~Acta}}   
\def\grl{\aaref@jnl{Geophys.~Res.~Lett.}} 
\def\jcp{\aaref@jnl{J.~Chem.~Phys.}}      
\def\jgr{\aaref@jnl{J.~Geophys.~Res.}}    
\def\jqsrt{\aaref@jnl{J.~Quant.~Spec.~Radiat.~Transf.}}
\def\memsai{\aaref@jnl{Mem.~Soc.~Astron.~Italiana}}
\def\nphysa{\aaref@jnl{Nucl.~Phys.~A}}   
\def\physrep{\aaref@jnl{Phys.~Rep.}}   
\def\physscr{\aaref@jnl{Phys.~Scr}}   
\def\planss{\aaref@jnl{Planet.~Space~Sci.}}   
\def\procspie{\aaref@jnl{Proc.~SPIE}}   

\begin{document}

  \title{\textsl{INTEGRAL} observation of the accreting pulsar GX~1+4}

   \author{Carlo Ferrigno
          \inst{1}
          \and
          Alberto Segreto\inst{1} %
      \and
      Andrea Santangelo\inst{2}
      \and
      J\"orn Wilms\inst{3,6}
      \and
      Ingo Kreykenbohm\inst{2,4}
      \and
      Miroslav Denis\inst{5}
      \and
      R\"udiger Staubert\inst{2}
          }

   \authorrunning{C. Ferrigno et al}
   \titlerunning{\textsl{INTEGRAL} observation of GX 1+4}
   \offprints{C. Ferrigno}

   \institute{IASF-INAF, via Ugo la Malfa 153, 90136 Palermo Italy\\
              \email{ferrigno@ifc.inaf.it}
         \and
        IAAT, Abt.\ Astronomie, Universit\"at T\"ubingen,
                Sand 1, 72076 T\"ubingen, Germany
     \and
        Department of Physics, University of Warwick, Coventry
              CV4 7AL, United Kingdom
    \and
        \textsl{INTEGRAL} Science Data Centre, 16 ch. d'\'Ecogia, 1290 Versoix, Switzerland
     \and
        Space Research Center, Bartycka 18a, 00716 Warsaw, Poland
	\and
		Dr. Remeis Sternwarte, Astronomisches Institut University of Erlangen-Nuremberg, Sternwartstr. 7 96049 Bamberg, Germany
             }

   \date{Received ---; accepted ---}

   \abstract{
    We present the results of the \textsl{INTEGRAL}
    monitoring campaign on the accreting low mass X-ray binary pulsar GX~1+4
    performed during the Galactic plane scan of the \textsl{INTEGRAL} Core Programme.
    The source was observed in different luminosity states ranging from
    $L_{20-40\,\mathrm{keV}}= 1.7 \times 10^{-10}\mathrm{erg\,cm^{-2}\,s^{-1}}$,
    to $L_{20-40\,\mathrm{keV}}= 10.5 \times 10^{-10}\mathrm{erg\,cm^{-2}\,s^{-1}}$
    for about 779\,ks from March 2003 until October 2004.
    Our observations confirm the secular spin down of GX~1+4
    with the spin period ($P_\mathrm{s}$) varying from 139.63\,s to 141.56\,s.
    In the
    highest luminosity state, a spin-up phase is observed.
    The phase-averaged spectrum of the source was modelled
    either with an absorbed cut-off power law or with a Comptonization
    model with significantly different parameters in the two
    brightest luminosity states. No evidence of any
    absorption-like feature
    is observed in the phase averaged spectrum up to 110\,keV.
    At highest luminosity, the source is found to pulsate up to 130\,keV. Phase resolved spectroscopy reveals
    a phase-dependent continuum and
    marginal evidence for an absorption feature at $34\pm2$\,keV in the descending
    part of the pulse.
    If interpreted as due to electron resonant cyclotron scattering, the magnetic field in the emitting
    region would be $(2.9\pm0.2) \times 10^{12} (1+z)$\,G where $z$ is the gravitational red shift
    of the emitting region.
    We also observed a very low luminosity state,
    typical of this source, which lasted for about two
    days during which the source spectrum was modelled by a
    simple power law, and a pulsed signal was still detectable in the 15--100\,keV energy range.
     }

   \maketitle

\section{Introduction}

The X--ray binary pulsars (XRBPs) were discovered more than 30 years ago with the pioneer
observation of Cen~X-3 by \citet{giac71}.
Even though the basic mechanisms of such a pulsed emission were
understood quickly \citep{Pringle72,Davids73}, XRBPs still present many puzzling aspects. In
particular, their wide-band spectral behaviour has not been
explained yet on the basis of a unique coherent physical model.
Radiation from XRBPs originates from the accretion of ionized gas,
from a nearby companion, an O or B star ($M \geq 5M_{\odot}$) in
High Mass X-ray Binaries (HMXRBs), a later than type A star ($M\leq M_{\odot}$)
or a white dwarf in low mass X-ray binaries (LMXRBs),
onto the magnetic pole regions of strongly magnetized (B
$\sim$10$^{12}$ G) rotating neutron stars. At several hundred
neutron star radii, the plasma is threaded and then funneled
along the magnetic field lines onto the magnetic
poles at the neutron star surface \citep{Pringle72,Davids73}.
Pulsation is generated because of the
nonalignment between the magnetic and rotational axes. Spin periods
are not constant and  \emph{spin--up} and \emph{spin--down}
behaviours are observed on a
variety of time scales
\citep{bildsten1997}.

According to accretion theory
\citep{ghosh1977,ghosh0,ghosh1979}, the accretion process onto a
neutron star surrounded by a disk strongly depends on the structure of the transition
zone where the Keplerian disk meets the pulsar magnetosphere. The
occurrence of accretion is mainly governed by the ratio between
the magnetospheric radius $r_m$, where the ram pressure of a
spherically symmetric inflow equals the magnetic pressure, and the
corotation radius $r_{co}$, where the magnetic field lines move at
the local Keplerian velocity.

When $r_\text{m} < r_\text{co}$, the plasma is forced to stream
along the field lines and eventually falls onto the neutron star's magnetic poles.
When $r_\text{m} > r_\text{co}$, the accretion stops due to the onset of the centrifugal
barrier, which is known as the {\it propeller effect} \citep{illarionov1975,stella1986}.
If $r_\text{m} \sim r_\text{co}$, the system is
in a near equilibrium state. In this condition,
\emph{spin--up} and \emph{spin--down} periods, due to the transfer of angular
momentum from the accreted plasma to the NS, can occur when variations in the
accretion rate ($\dot M$) take place.
When $\dot M$ is sufficiently high, the star experiences a spin--up torque; as
$\dot M$ decreases, the spin--up torque falls and vanishes at a critical accretion rate.
For a lower accretion rate, the star experiences a spin--down torque whose magnitude
increases until $\dot M$ reaches the minimum accretion rate consistent with
steady accretion \citep{ghosh1979}.

Such torque reversal episodes characterise the
peculiar spin history of the ($P \sim 140$\,sec)
LMXRB GX~1+4. The source, which
appears to accrete from the wind of the 18 Mag, M5III
companion, V2116~Ophiuchi \citep{davidsen1976},
exhibited in the 1970s the highest spin
up-trend ever observed in any pulsar with $\dot{\nu}/\nu$: as high as
$\sim$2\% per year \citep{nagase}.
The orbital period of the source was reported to be $\sim 304$~days by \citet{pereira1999}.
More recently,
\citet{hinkle2003} reported a $P_{\mathrm{orb}}\sim 1042$~days, based on infrared
spectroscopic properties of V2116~Ophiuchi.

No observations were made from 1980 to 1982; later
EXOSAT observations in 1983 and 1984 failed to detect the source, indicating an
X-ray flux decrease of at least two orders of magnitude
\citep{hall1983}. In 1986, GX~1+4 was observed during a balloon flight
\citep{greenhill1989} and exhibited a spin period of 111.8\,s,
indicating that one or more spin--down episodes occurred between 1980
and 1986. However, this episode must have been localized in time, since a
collection of different data since 1980 shows that GX~1+4 was again spinning
up at an average rate of $6.6\times 10^{-8}$\,s/s
similar to the rate it had in the 1970s \citep{greenhill1989}.

The pulsar reappeared with a reversed period derivative and low
luminosity in 1987 \citep{makishima1988,mony1991}; and during
1989--1991, it continued to spin down rapidly with a mean rate about
half, in absolute value, of the previous spin--up rate
\citep{chakba1997}.

On an irregular basis, GX~1+4
presents spin-up episodes with the absolute intensity similar to the spin-down
trend. To explain such episodes
on the basis of some apparent anti-correlation between torque and luminosity
in the 20--60\,keV band %
seen by BATSE, several authors \citep{chakba1997,nelson1997}
have suggested the formation of a retrograde disk
implying a magnetic field of $B\sim10^{12}$\,G on the surface of the neutron star.
Instead, other authors \citep{beurle1984,dotani1989,mony1991}
inferred the highest dipole magnetic
field of any accreting pulsar from the torque reversal episodes: $B = (1$--$3)\times
10^{13}$\,G on the surface of the neutron star.

Cyclotron resonant scattering features (CRSFs) provide
a tool for direct measurement of the
magnetic-field strengths of accreting pulsars since
they appear at
$E_{cyc} = 11.6\,B_{12}\times (1+z)^{-1}$\,keV, where $B_{12}$ is
the magnetic field strength in units of $10^{12}$\,G and $z$ is
the gravitational red shift of the emitting region \citep{coburn2002, disalvo2004}.
No CRSFs have been detected until today in the spectrum of GX~1+4.
This lack of detection is not surprising if the estimate based on the standard accretion torque theory
is correct. In fact, the
inferred magnetic field of $\sim 10^{13}$~G converts to
$E_\text{cyc}\approx 100$--$300$\,keV, which is beyond the detection limit of
hard X-ray satellites like \textsl{Beppo-SAX} and \textsl{RXTE} for the flux level and
spectral shape of GX~1+4 \citep{naik2004,galloway2000}.

Moreover, evidence has been found that the hardness of the XRBP
spectrum is closely related to the field intensity.
\textsl{Ginga}, \textsl{Beppo-SAX}, and \textsl{RXTE} observations of binary accreting pulsars have
revealed a correlation between line
energy and spectral hardness in the high-energy part of the spectrum
\citep{makishima1992,dalfiume2000,coburn2002}.
Since the observed spectrum of GX~1+4 is one of the hardest found among all X--ray binary pulsars,
this relation suggests a magnetic field in excess of $10^{13}$\,G. Similar results have been found
for GS~1843-00 \citep{piraino2000}.

In intermediate and high luminosity states
($L_{2-10\,\mathrm{keV}} = 10^{-10}$--$10^{-9}\, \mathrm{erg\,cm^{-2}\,s^{-1}}$),
its emission has been described either with a cut-off power law
\citep{kotani1999,cui1997,cui2004,rea}
or with a Comptonization model \citep{galloway2000,galloway2001,naik2004}.
In both cases the source shows high intrinsic absorption
and exhibits a strong Iron line at 6.5--7\,keV with an equivalent width of 0.2--0.5\,keV.

On an irregular basis, GX~1+4 presents low luminosity states in which the flux
decreases to a few $L_{2-10\,\mathrm{keV}} = 10^{-11}\,\mathrm{erg\,cm^{-2}\,s^{-1}}$.
In these states, even though the other spectral characteristics are highly variable,
the source always shows an enhanced Fe emission at 6.5--7\,keV
(few keV equivalent width) and a harder spectrum in which the
cut-off often cannot be found or is at higher energy with respect
to the bright state \citep{rea,naik2004,cui1997,cui2004}. These
episodes can last from about one day
\citep{galloway2000low,rea,naik2004} to several months
\citep{cui1997,cui2004}. The phenomenological interpretation is
ambiguous: \citet{cui2004} explain the low luminosity
episodes as due to the propeller effect, while the residual power
law emission is produced by accelerated particles;
\citet{galloway2000low} present a model involving the accretion
column disruption; \citet{rea} suggest that the pulsar is hidden
by a thicker accretion disk. In the last case, the residual
radiation is produced by the reflection of the disk edge and by
partial transparency of the medium at high energies ($\ga
10$\,keV).

The main goal of this paper is to study the spectral and
timing characteristics of GX~1+4 observed over a wide range of
luminosity, within the \textsl{INTEGRAL}
Galactic Plane Scan (GPS)
part of the Core Programme \citep{gps} from March 2003 until
April 2004. The search for CRSF is also one of the key goals of
our study.

In Sect.~\ref{sec:observation} we describe the observations and
the data analysis method. In
Sect.~\ref{sec:results} we report on the results obtained from analysing
light curves, spin period, pulse profiles and phase-resolved and
phase-averaged spectra. In Sect.~\ref{sec:discussion} we discuss
our results in relation to previous observations. Finally,
in Sect.~\ref{sec:conclusions} we draw our conclusions.

\section{Observations and data analysis.}
\label{sec:observation}
The European Space Agency's International Gamma-Ray Astronomy
Laboratory (\textsl{INTEGRAL}), launched in October 2002, carries 3 co-aligned
coded mask telescopes.
\begin{itemize}
\item The imager \textsl{IBIS} \citep[Imager on Board the \textsl{INTEGRAL} Satellite;][]{ibis}, which allows for
12\,arcmin angular-resolution imaging in the energy range from 15\,keV to
600\,keV with energy resolution of 6--7\%;
IBIS is made of a low-energy (15--600\,keV) CdTe detector \textsl{ISGRI}
\citep{isgri} and of a CsI layer, \textsl{PICsIT} \citep{picsit}, designed for optimal performance at
511\,keV, with an overall nominal energy range between 175\,keV and 10\,MeV.
\item The spectrometer \textsl{SPI} \citep[Spectrometer on \textsl{INTEGRAL};][]{spi},
sensitive from 20\,keV to 8\,MeV with an angular resolution
of $2.5\deg$ and an energy resolution of a few times $10^{-3}$.
\item The X-ray monitor \textsl{JEM-X} \citep[Joint European X-ray Monitor;][]{jemx} made of
two independent units \textsl{JEM-X1} and \textsl{JEM-X2}, sensitive from 3\,keV to 34\,keV, with an
angular resolution of 1\,arcmin and an energy resolution of 10--15\%.
\end{itemize}

\begin{figure}
  \begin{center}
    \resizebox{\hsize}{!}{\includegraphics[angle=0]{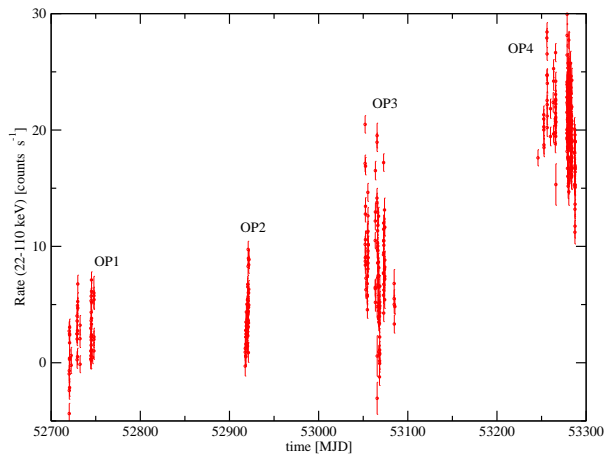} }
    \caption{Light curve of GX~1+4.
    We plot the \textsl{INTEGRAL} \textsl{ISGRI} count rates in the 22--110\,keV band
    during the galactic plane scan pointings with an off-set angle less than $6\deg$.
    The binning is one science window corresponding to 1.6--3.6\,ks.
    The time scale is modified Julian days (MJD=JD-2\,400\,000.5); MJD 52700 is 1 March 2003.}
 \label{fig:lc_mjd}
\end{center}
\end{figure}

As part of its core programme, \textsl{INTEGRAL} performs regular scans
of the Galactic plane (GPS),
with the aim of monitoring the timing and spectral properties of the known
X-ray sources, discovering new transient sources, and mapping
the diffuse emission of the Galactic plane \citep{gps}.

\begin{table*}
\begin{center}
\caption{
    Summary of the \textsl{INTEGRAL} observations of GX~1+4.
    In the table we show
    the name for the observational period, its start and end time,
    the exposures of the three \textsl{INTEGRAL} instruments used here
    and the corresponding total number of source counts.
}

\begin{tabular}{ l l l c c c c c c  }
\hline
\hline
name & start [MJD] & end [MJD] &   \multicolumn{3}{c}{exposure[ks]} & \multicolumn{3}{c}{Source counts} \\\
           &                  &                   & \textsl{ISGRI} & \textsl{JEM-X} & \textsl{SPI} & \textsl{ISGRI} & \textsl{JEM-X} & \textsl{SPI}\\
\hline
OP1  & 52\,720.8076 & 52\,748.9027  &  72 & & & $2.9\times10^5$ & & \\
OP2  & 52\,917.6381 & 52\,921.7215  &  97 & & & $5.2\times10^5$ & & \\
OP3  & 53\,052.0923 & 53\,085.5729  & 155 & 47 & 137 & $1.9\times 10^6$ & $2.0 \times 10^5$ & $4.2 \times 10^3$ \\
OP4  & 53\,252.4430 & 53\,287.3416  & 423 & 154& 351 & $2.4\times 10^6$ & $3.8 \times 10^5$ & $3.9 \times 10^3$ \\
OP3L & 53\,067.2531 & 53\,068.7772  &  32 & & & $9\times10^4$ & & \\
\hline
\label{tab:obs}
\end{tabular}
\end{center}
\end{table*}

From March 21, 2003 to September 17, 2004 the region containing GX~1+4
was observed several times. In order to have the
shadowgram of GX~1+4 fully coded
on the \textsl{ISGRI} detector plane,
only pointings where GX~1+4 is observed at less than $6\deg$
off-axis were used.
This gives the best S/N performance for ISGRI.

The data analysed in this paper consist of 435 science windows for a total observing time of 779\,ks
and are divided into four intervals of semi-contiguous science windows as
shown in the light curve of Fig.~\ref{fig:lc_mjd}.
We define time intervals as follows: MJD 52\,720--52\,750 observational period 1 (OP1),
MJD 52\,915--52\,925 (OP2), MJD 53\,050--53\,090 (OP3), and MJD 53\,250--53\,290 (OP4).
We also distinguish
one sub-period corresponding to a low-flux episode in the middle of an
intermediate flux state (OP3L), which started on MJD 53\,067.25 and ended on
MJD 53\,068.75.
In OP1, OP2, and OP3L, the source flux was too low for a detection with \textsl{JEM-X} or \textsl{SPI}.
In OP4 and in the last part of OP3 (after MJD 53\,072), it was possible to use only \textsl{JEM--X1},
while in the first part of OP3 only \textsl{JEM--X2} was available.
In OP3 we did not analyse the 14\,ks of \textsl{JEM--X1} data due to the low
statistics; the reported exposure
refers to \textsl{JEM--X2}. The lower exposure time of \textsl{JEM--X} compared to \textsl{ISGRI} is due to
the smaller field of view of this instrument.
The complete list of the observations is reported in Table~\ref{tab:obs}.

We used version 5.1
 of the \textsl{INTEGRAL} off-line standard analysis software (OSA~5.1),
 officially distributed by the \textsl{INTEGRAL} Science Data
Centre (ISDC), to obtain the phase-averaged spectra of SPI and
\textsl{JEM-X}. For the latter instrument we limited the analysis to the energy range
4--21\,keV, for which the \textsl{JEM-X} team ensures a reliable calibration
as stated in the OSA handbook. All the information regarding
the official software and the standard calibration can be found on the ISDC
web-site at\\
\texttt{http://isdc.unige.ch/}.

To analyse the \textsl{ISGRI} data we had to face two
kinds of problems that are not optimally solved by the OSA pipeline: the
lack of tools for performing straightforward timing and phase-resolved
analysis and a not-yet-optimized energy reconstruction of the
events. This required the development of specific tools, which are described in the appendix.

\section{Results}
\label{sec:results}
\subsection{Light curves}
The ISGRI light curve of GX~1+4 in the 22--110\,keV energy range,
averaged over the science window exposure (typically 1.6-3.6\,ks), is
shown in Fig.~\ref{fig:lc_mjd}. During the \textsl{INTEGRAL} AO1
Core Program Monitoring, GX~1+4 evolved from a weak state
($L_{20-40\,\mathrm{keV}} \simeq 2 \times
10^{-10}\mathrm{erg\,cm^{-2}\,s^{-1}}$) to a much brighter one
($L_{20-40\,\mathrm{keV}} \simeq 10 \times
10^{-10}\mathrm{erg\,cm^{-2}\,s^{-1}}$).

\begin{figure}
  \begin{center}
    \resizebox{\hsize}{!}{\includegraphics[angle=0]{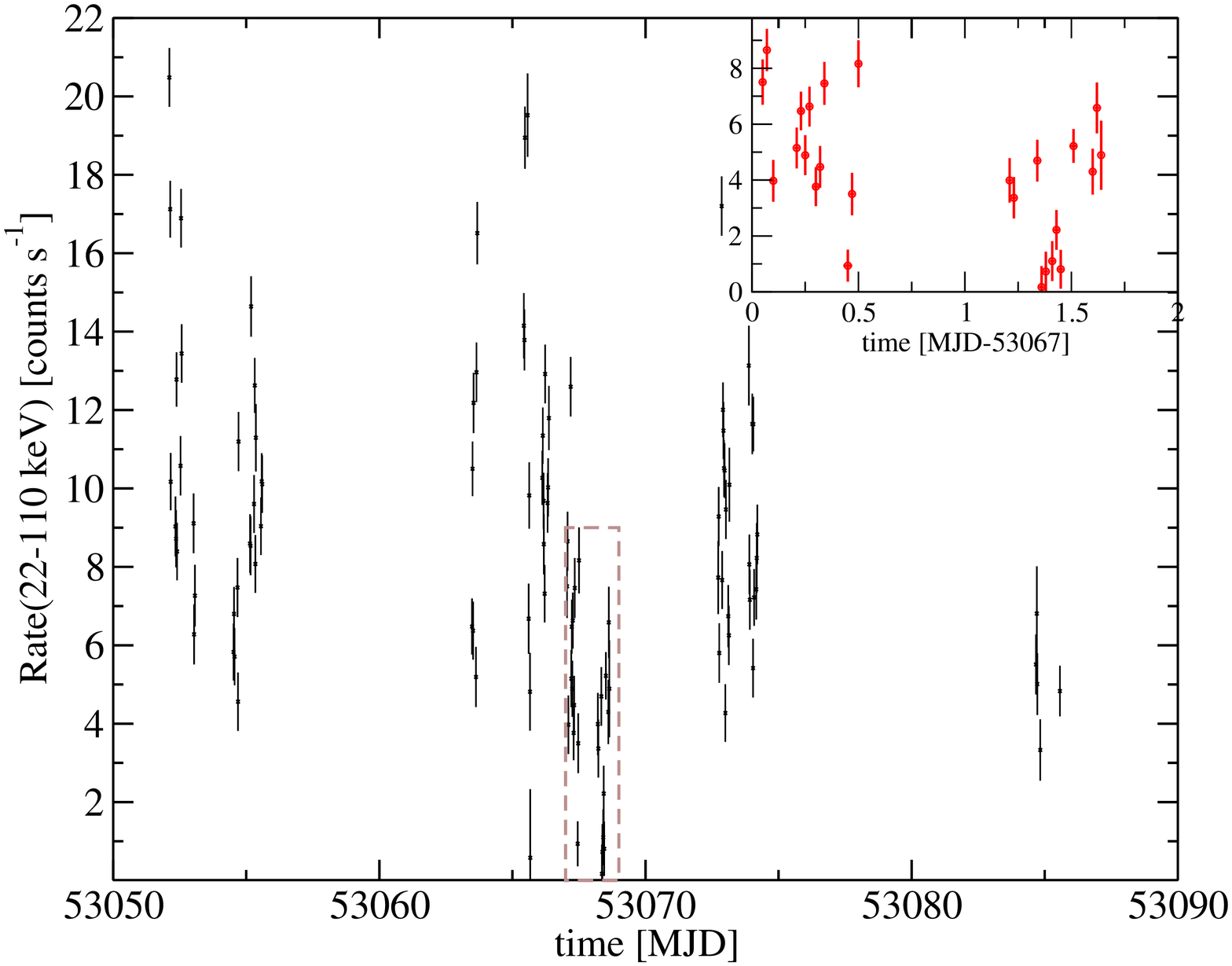} }
    \caption{Light curve of GX~1+4 observed by \textsl{ISGRI} during the
    GPS for OP3 in the 22--110\,keV energy band.
    The region with the dashed border highlights the low-luminosity
    episode, which is reproduced in the inset
    in the upper right corner. The binning is one science window
    corresponding to about 1.7--3.6\,ks depending on the operational
    mode. Errors are at $1\sigma$ level.}
\label{fig:lc_mjd_164}
\end{center}
\end{figure}

During OP3, from February 17, 2004 to March 23, 2004, the source
exhibited erratic variations of
about one order of magnitude in flux on a few kilo-seconds time scale
(see Fig~\ref{fig:lc_mjd_164}).
In particular, from MJD~53\,067.25 until MJD~53\,068.75,
GX~1+4 underwent a very low luminosity state (OP3L) observed by \textsl{INTEGRAL}
for $\sim$32\,ks.

Since these low-luminosity episodes
are peculiar in
the behaviour of the source \citep{rea,galloway2000,cui2004}, OP3L data are analysed
separately and excluded from the average analysis
of OP3.

\subsection{Spin period}
\begin{figure*}
  \begin{center}
   \includegraphics[angle=0,width=8.4cm]{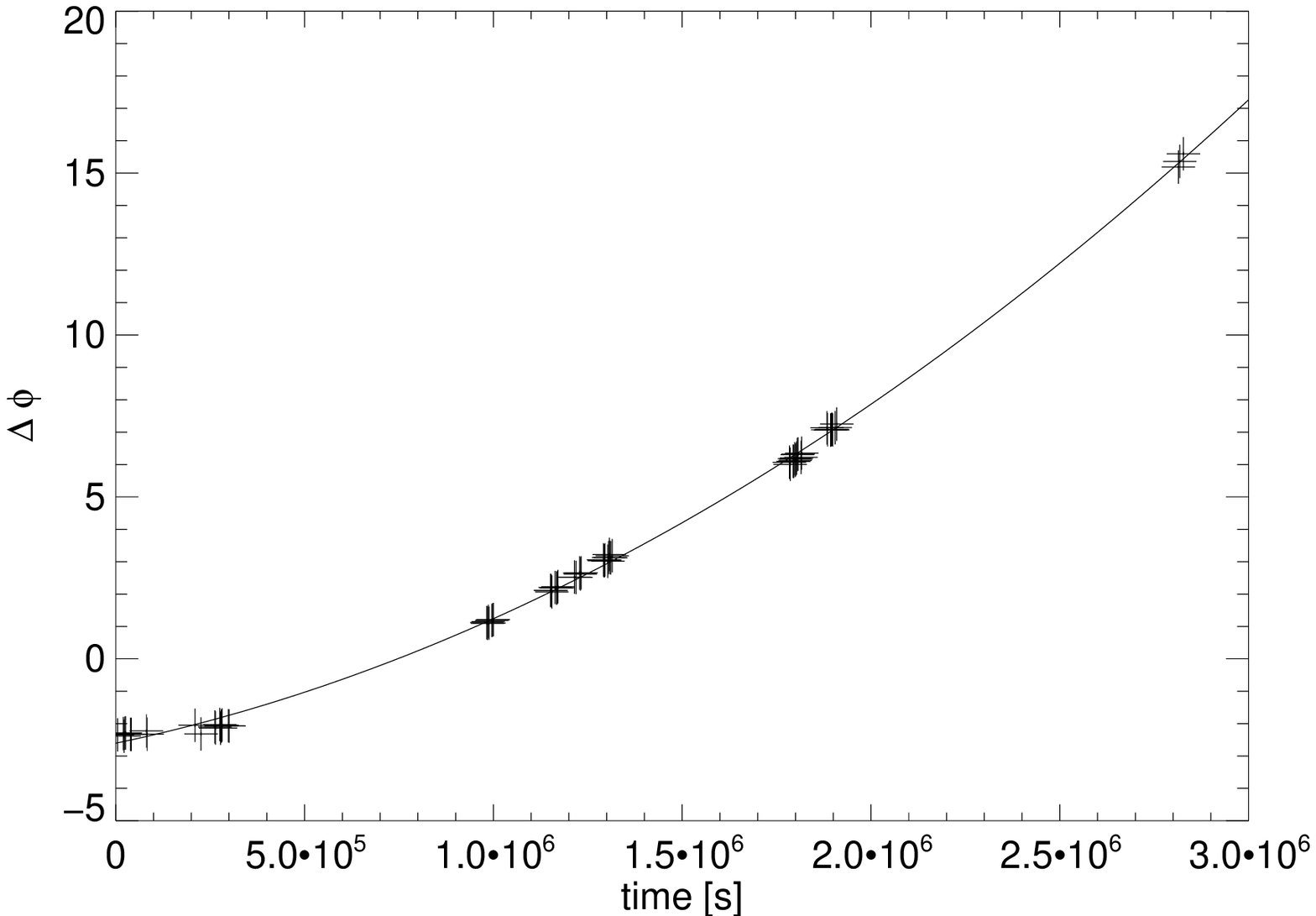} \hfill
   \includegraphics[angle=0,width=8.4cm]{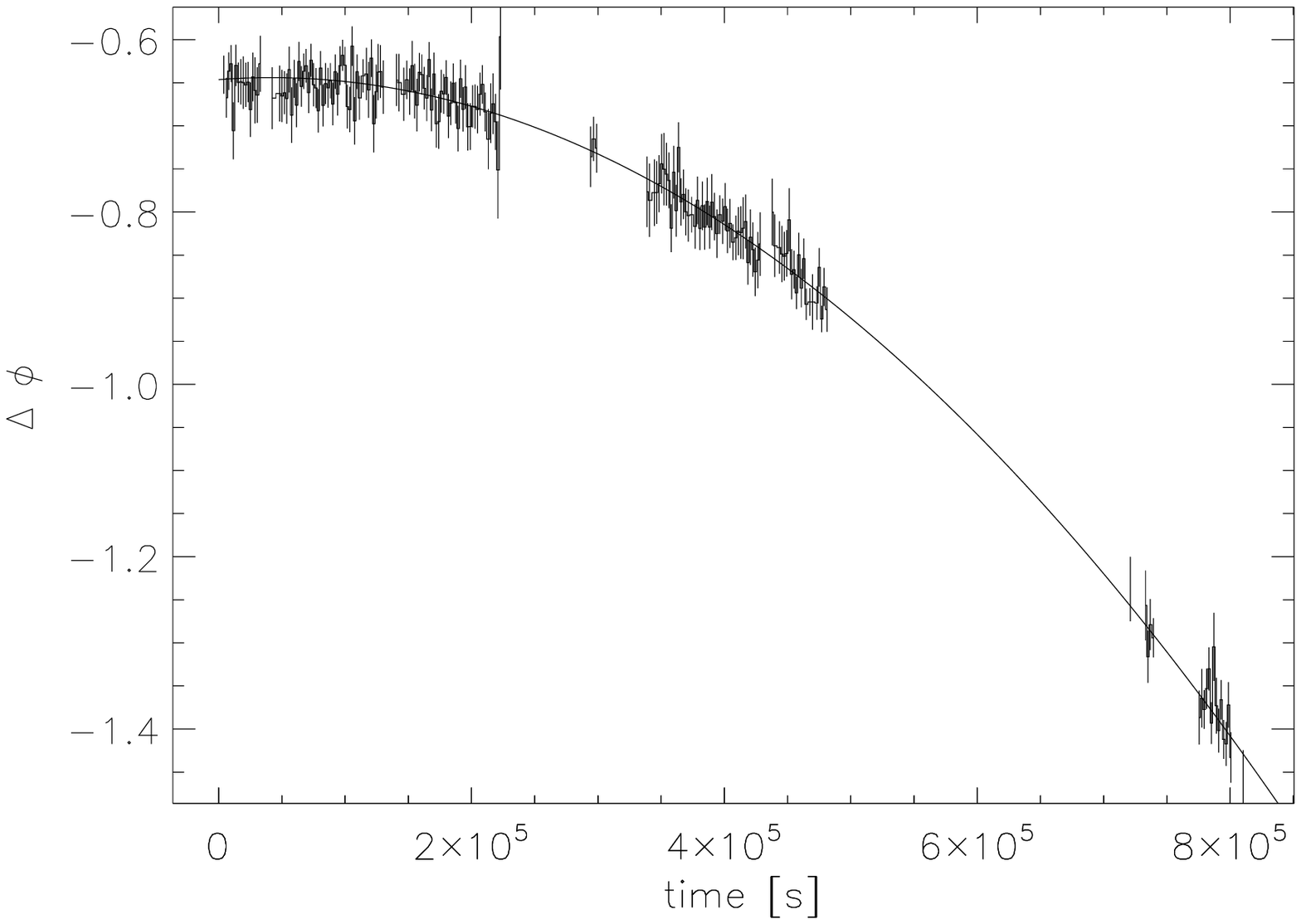}\\
   \includegraphics[angle=0,width=8.4cm]{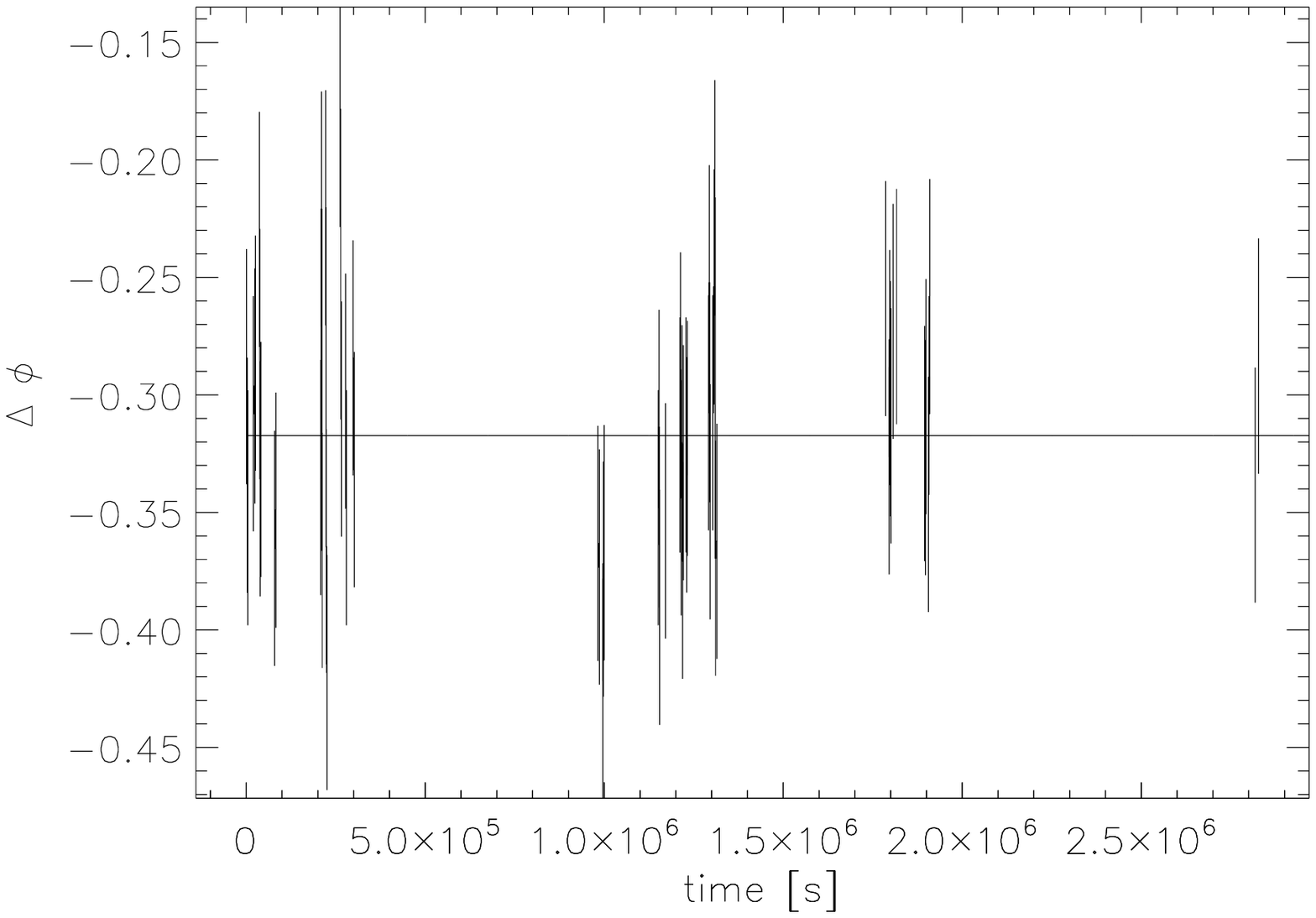} \hfill
   \includegraphics[angle=0,width=8.4cm]{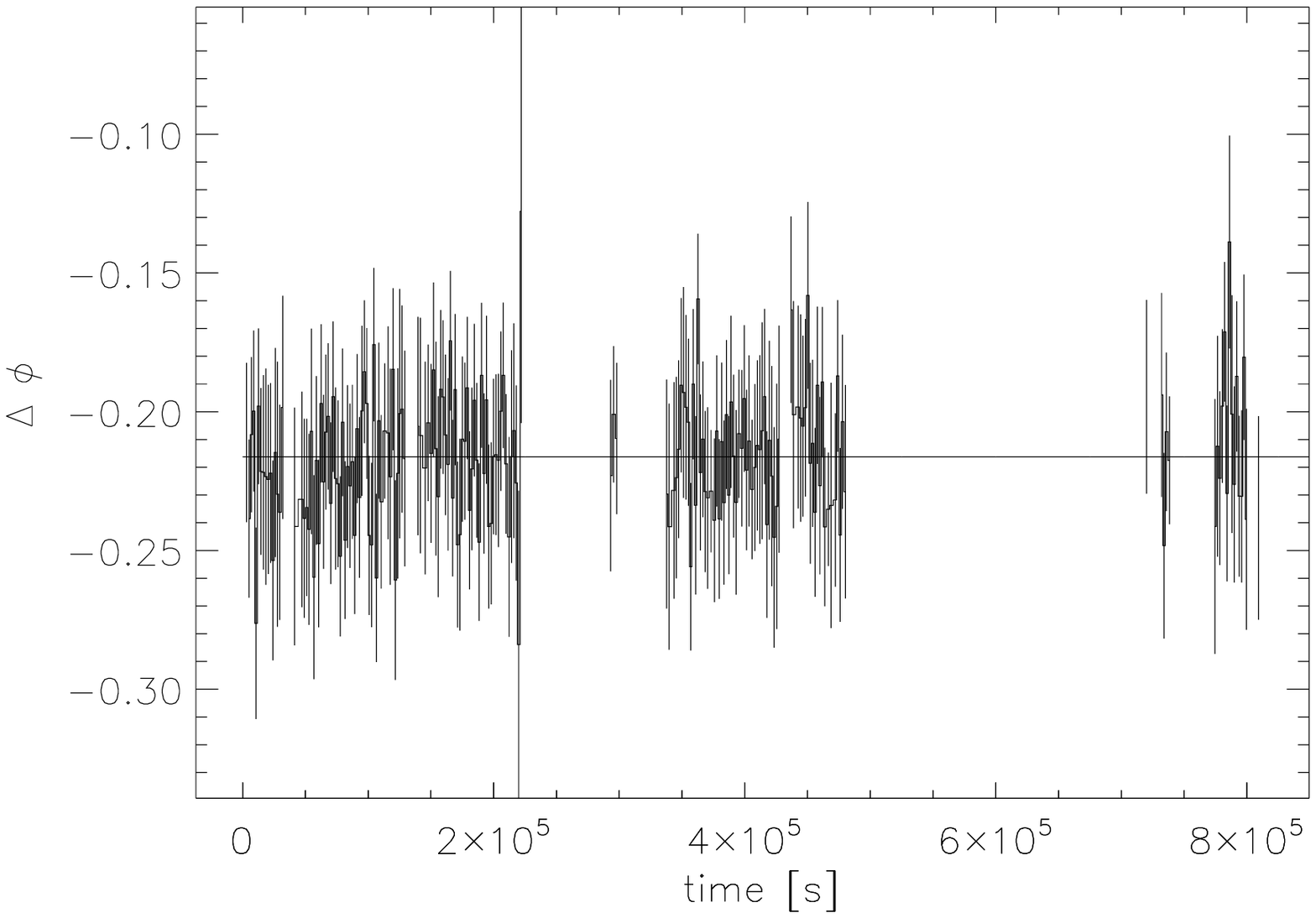}
    \caption{
    The phase shifts are obtained using a constant folding period with
    the best-fit second-order polynomial overplotted (upper panels) and the residual phase shifts obtained with
    the folding parameters of Table~\ref{tab:spin_period} (lower panels).
    The phase shifts can be converted into time units by multiplying by the folding period of 140\,s.
    Left is for OP3 and right for OP4. The time scale is in seconds after
    the reference times of Table~\ref{tab:spin_period}.}
\label{fig:delta_phi}
\end{center}
\end{figure*}

Spin frequencies in the four observational periods were determined
separately.
First, we computed a preliminary power spectrum from the
light curve binned with a 10\,s time resolution. The $\sim 0.007$~Hz peak position
gave us a tentative period from which we
produced a pulse profile $P_t(\phi)$ for each science window.
Then, we computed the phase of the maximum of the
pulse first Fourier component in each science window:
\begin{equation}
\phi_0(t) = \tan^{-1}( I_c(t)/I_s(t))\;,
\end{equation}
where
\begin{equation}
I_c(t) = \int P_t(\phi)  \cos(\phi) \text{d}\phi \quad \text{and} \quad
\end{equation}
\begin{equation}
I_s(t) = \int  P_t(\phi)  \sin(\phi) \text{d}\phi \;.
\end{equation}
and $t$ is the science window central time.\footnote{For the pulse shape of GX~1+4, most
of the power is contained in the first harmonic; for sources with a double peak
this method should be used with caution.}
If the law we used for folding is not optimal, $\phi_0(t)$
is not constant throughout the observational period; however, we can derive the optimal folding law from its
variation.
Let $\Delta\phi_f(t) = \phi_0(t)-\phi_0(t_0)$ be
the phase shift measured after folding the data with the temptative period; we reconstruct the
actual phase shift $\Delta\phi(t)$
by fitting $\Delta\phi(t)$ with a second-order polynomial
\begin{equation}
\Delta \phi (t) = \phi_0 + a (t-t_0) + b (t-t_0)^2 \, ,
\end{equation}
where $\phi_0$ is the starting phase shift determined by the fit
and $t_0$ the reference folding time chosen close to the beginning of
the observational period. The coefficient $a$ is used to correct
the tentative choice of the folding frequency $\nu_t$ as
\begin{equation}
\nu_c = \nu_t - a \, ,
\end{equation}
where $\nu_c$ is the corrected pulse frequency. The coefficient $b$ is linked to the
pulse frequency derivative during the observational period by the relation
\begin{equation}
\dot \nu_c = -2 b \, .
\end{equation}
Finally, we verified that $\Delta\phi(t)$ obtained with the
corrected folding law is constant within the errors and therefore the derived parameters
are correct. A visual inspection of the residuals in Fig.~\ref{fig:delta_phi}
confirms that it is not necessary to use a higher-order polynomial and that
the connection of phase shifts separated by wide gaps within the OPs is self
consistent, even though in principle it could be non-unique.
The different OPs cannot be phase-connected due to their time separation
and to the complex spin evolution of the source, with e.g. a torque reversal between OP3 and OP4.

\begin{table*}
\caption{ Values for the spin period ($P$) and spin period
derivative ($\dot P$), in the three observational intervals where
the source was bright enough for this determination, are reported.
The errors are at 1$\sigma$ level computed
from the fit to the phase shifts.
}
\begin{center}
\begin{tabular}{ l l c c c }
\hline
\hline
conv. name & reference time [MJD] & $P$ [s] & $\dot P $[s/s] \\
\hline
OP2 & 52917.638038  & $139.630\pm0.006$    & $(1.5\pm 0.1)\times 10^{-7}$     \\
OP3 & 53052.091969  & $140.6132\pm0.0002$  & $(1.034\pm 0.002)\times 10^{-7}$  \\
OP4 & 53252.442934  & $141.56488\pm0.00014$& $(-0.5331\pm 0.0004)\times 10^{-7}$\\
\hline
\end{tabular}
\end{center}
\label{tab:spin_period}
\end{table*}

Due to the low number of source counts for OP1, spin periods and their derivatives
were only measured for OP2, OP3, and OP4. In Table
\ref{tab:spin_period} we summarise the values for the spin
period and its first derivative, and the reference time at which they are measured.
The relative time intervals are specified
in Table~\ref{tab:obs}. In some science windows in OP2
and OP3, the source was too weak for a good determination of the
phase of the maximum, we so excluded them from the analysis, which
produced an exposure time reduction of less than 5\%. The source shows the longest spin
period of its observational history $P_\mathrm{s}\sim$141\,s, which
confirms the overall spin down trend \citep{chakba1997}. In OP4 we
find a sign inversion of the spin period derivative, which is the
signature of a torque reversal episode.

The secular spin-down trend we measured from a linear fit to the three measured pulse period values
is $\dot P~=~(6.6\pm0.8)\times10^{-8}$\,s/s, but from the analysis of phase shifts we find
local spin period derivatives that are larger in absolute value. This is consistent with
the source behaviour in previous observations \citep{chakba1997,pereira1999}: local spin-period
variations are superimposed on a global spin down trend.

\begin{figure*}
  \begin{center}
    \resizebox{!}{19cm}{\includegraphics[angle=0,height=19cm]{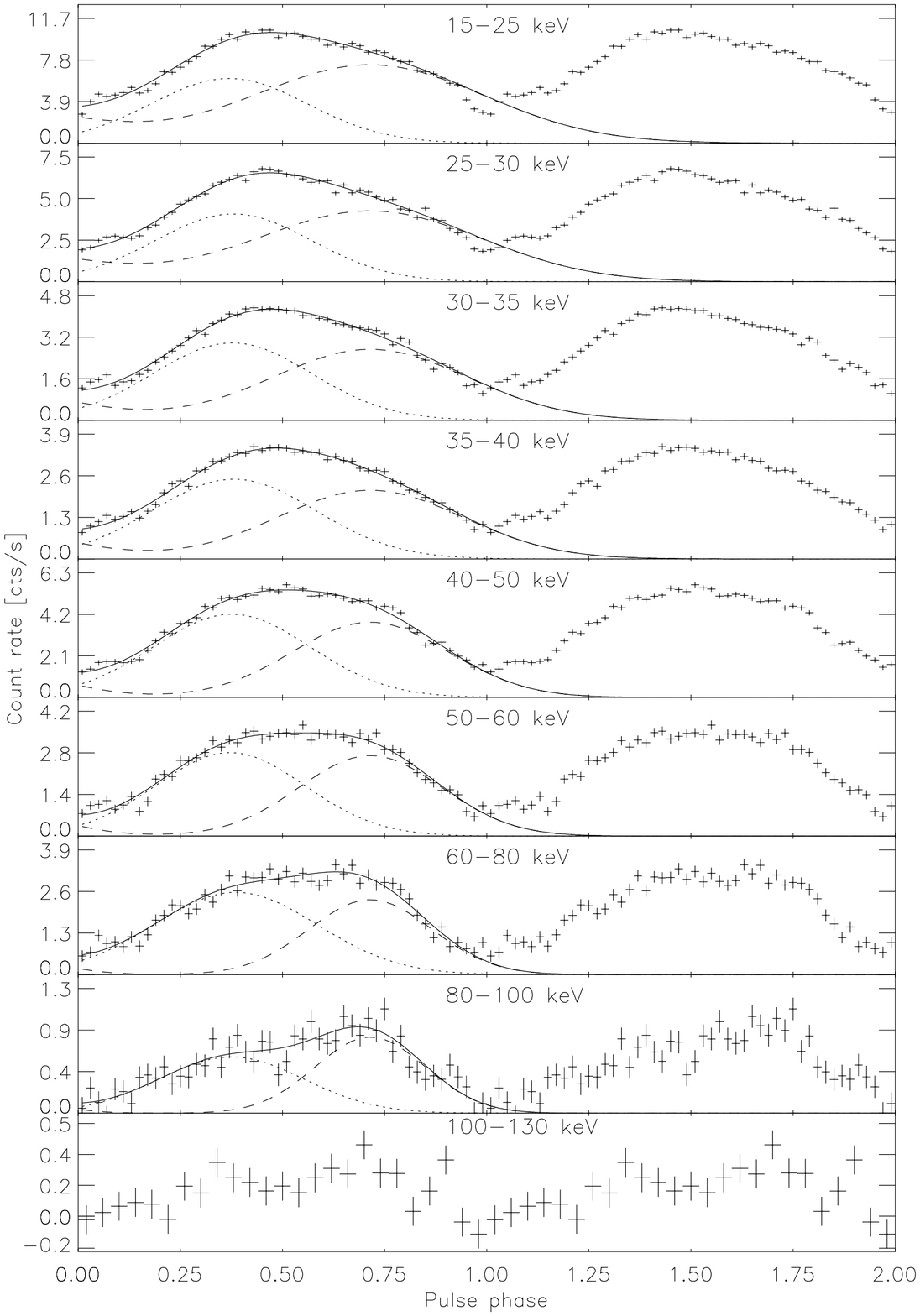}}
    \caption{The background subtracted pulse profiles of OP4 with energy ranges indicated in the
    figure. The vertical scale is in counts per second, the reference time, spin period, and
    spin-period derivative are listed in Table~\ref{tab:spin_period}. The solid lines are the best-fit pulse profiles, the dotted and the dashed lines
    are the two Gaussian subcomponents used in the fit.}
\label{fig:op4_pulses}
\end{center}
\end{figure*}

\begin{figure}
  \begin{center}
    \resizebox{\hsize}{!}{\includegraphics[angle=0]{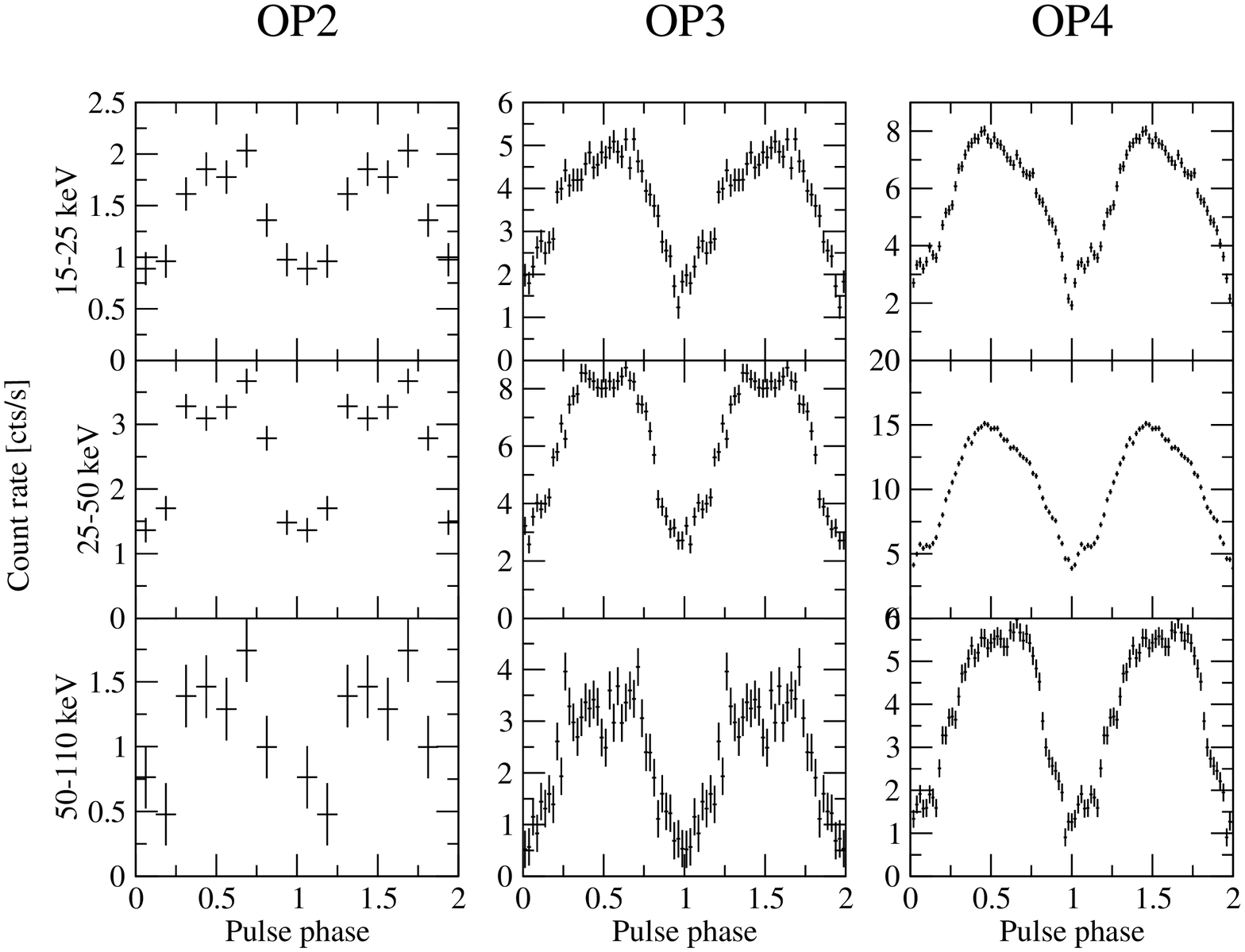}}
    \caption{The background subtracted pulse profiles for OP2, OP3, and OP4
    in the energy ranges indicated in the plot.
    The reference time, spin period, and
    spin period derivative are listed in Table~\ref{tab:spin_period}.
    }
\label{fig:pulses}
\end{center}
\end{figure}

\subsection{Pulse profiles}
Using the values of $P$ and $\dot P$ in Table~\ref{tab:spin_period}, we
extracted the background subtracted pulse profiles of \textsl{ISGRI} for different
energy intervals using the tools described in the appendix B. The pulse profiles
for OP4 are shown in Fig.~\ref{fig:op4_pulses}, the ones for OP2, OP3, and OP4 in
Fig.~\ref{fig:pulses}. Phase zero was arbitrarily chosen at the bottom of the
dip in the 15--25\,keV profile for OP4 and at the pulse minimum for the other
OPs. At the highest luminosity, the source is observed to pulsate up to 130\,keV.

As a first approximation all pulse profiles show a single broad peak --
as stated repeatedly in the literature. There are, however, clear substructures.
A dip like structure within the broad minimum (around pulse phase zero) is
evident at energies up to 60\,keV (Fig.~\ref{fig:op4_pulses}). This is a well-known
feature already found in \textsl{RXTE} and \textsl{Beppo-SAX} data
\citep[see e.g.][]{galloway2001,naik2004}.
The pulse appears asymmetric  %
and looks like a superposition of two (or more) peaks. This is also
strongly suggested by various pulse profiles found in the literature
\citep[see e.g.][]{giles2000,cui2004}. Following this suggestion, we
modelled our pulse profiles of OP4 by the sum of two Gaussian
components (leaving out four data points that define the dip around
phase 0).
The model fits are
shown as solid lines for all profiles below 100\,keV in Fig.~\ref{fig:op4_pulses}.
Each Gaussian is characterised by three parameters: centroid, width,
and amplitude (a constant flux was set to zero since it
was found to be unconstrained when it was left as a free parameter).
Most fits are quite good (reduced chi-squared $\sim$1), and others are
acceptable (reduced chi-squared up to 1.4 for 40 d.o.f.). Our
findings from these fits for OP4 (Fig.~\ref{fig:op4_pulses}) are the
following:

\begin{enumerate}
\item The centroids of both peaks are consistent with a constant
in phase throughout the entire energy range: 0.39 and 0.71, for the first and
second peaks, respectively.

\item The widths of both peaks are slightly variable: while for energies above
40\,keV both peaks have a width consistent with 0.18 phase units,
at lower energies the first peak becomes narrower
while the second peak becomes wider (reaching 0.16 and 0.26
phase units, respectively, for 15--25\,keV).

\item As evident from the visual inspection and from the ratio of
the amplitudes of both peaks, the second peak is the dominating one. The ratio
of amplitudes (first peak to second peak) increases from 0.5 at
15--20\,keV to 0.9 at 40--50\,keV with a subsequent small decrease to
$\sim$0.7 at 80--100\,keV.

\item There is only a weak indication of a third small peak inside the pulse
minimum around phase 0.1 (see, for instance, the 30--35\,keV profile),
reminiscent of the ``bump'' detected by \citet{greenhill1989}
and thereafter confirmed in other observations \citep[e.g.][]{naik2004,giles2000}.
\end{enumerate}

Pulse profiles of OP2 and OP3 (Fig.~\ref{fig:pulses})
confirm the general findings from OP4,
although in OP2 and OP3 the two subpeaks
contribute almost equally to the main pulse at all energies. An impressive
demonstration of the existence of different emission components in GX~1+4
and their variability is given in Figs.~4 and 5 of \citet{giles2000}
where three peaks (and the dip structure) are very apparent.

To obtain the pulsed fraction we fitted the pulse profiles with a sine function, plus a
constant and computed, from the sine amplitude $A$ and the constant $C$,
the pulsed fraction as $2A/(A+C)$. Pulsed
fractions in different energy intervals (where the S/N permitted
such an analysis) are reported in Fig.~\ref{fig:pulsed_frac}.
Interestingly, the intermediate luminosity observation (OP3)
shows an average higher pulsed fraction in comparison with the
brightest state (OP4). Pulsed fractions observed in OP2 are
consistent with the others due to the large errors in the
determination.

Using the spin period found for OP3, we extracted a pulse profile for the
low-luminosity episode (OP3L) in the 15--100\,keV energy band.
From the pulse profile of the right panel of Fig.~\ref{fig:veryweak}
we can infer a pulsed fraction of $0.59\pm0.14$ using the method outlined above.
\begin{figure}
  \begin{center}
    \resizebox{\hsize}{!}{\includegraphics[angle=0]{pulsed_fraction.eps}}
    \caption{The pulsed fraction in the three time intervals considered (see Table~\ref{tab:spin_period}).
    The crosses refer to OP2 (weak), the squares to OP3 (intermediate), and the
    circles to OP4 (high luminosity); the errors are $1\sigma$.
    }
\label{fig:pulsed_frac}
\end{center}
\end{figure}

\begin{figure*}
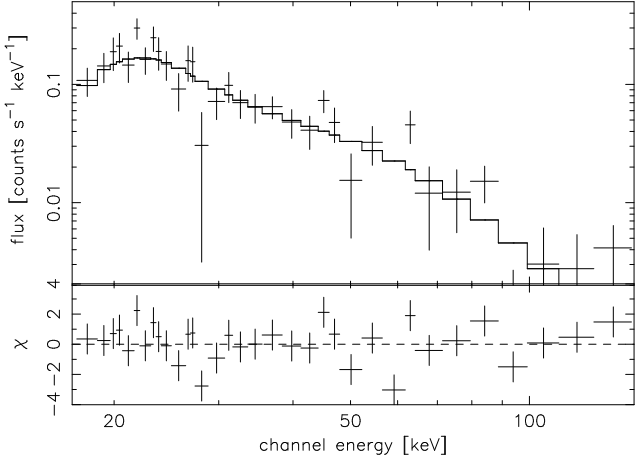

  \begin{center}
   \includegraphics[angle=270,width=8.4cm]{verylow.ps} \hfill
   \includegraphics[angle=270,width=8.4cm]{gx_verylow_pulse.ps}
    \caption{The \textsl{ISGRI} spectrum (left) and the pulse profile (right) in the low-luminosity episode OP3L.
    The spectrum extends from 17 to 150\,keV with no break and is fitted by a power law
    with a slope of $2.5\pm0.2$. The pulse profile is obtained in the 15--100\,keV energy band.
    }
\label{fig:veryweak}
\end{center}
\end{figure*}

\subsection{Spectral analysis}
\label{sec:ph_av_spec}
Spectral analysis was performed separately in the four OPs.
Results for phase-averaged spectra are reported here while in
Sect.~\ref{sec:phase_spectroscopy} we describe our results on phase-resolved spectroscopy.

Broad-band spectral analysis, combining data from \textsl{JEM-X},
\textsl{ISGRI}, and \textsl{SPI}, was feasible only for OP3 and OP4,
since the source was not detected by SPI and \textsl{JEM-X} in the
lower luminosity states. Thus, we firstly analysed the
\textsl{ISGRI} data alone for all luminosity states, and then we
extended our analysis to the other instruments for the higher
luminosity states.

The \textsl{ISGRI} spectra were fitted in the 17--110\,keV energy range using
a cut-off power law ($N(E)\propto\,E^{-\Gamma}\,\exp[-E/E_\text{c}]$) with the best-fit
parameters summarised in Table~\ref{tab:best_fit}. In OP1 and
OP2 the parameter determination is poor due to the weakness of the
source, so the fit results can be considered as generally consistent
with the others.
Instead, we note a significant difference between the best-fit parameters of OP3 and OP4,
which can be studied better by adding the data from \textsl{JEM-X} and \textsl{SPI}.

\begin{table}
\begin{center}
\caption{\textsl{ISGRI} spectral-fit parameters for the model $E^{-\Gamma}\,\exp(-E/E_\text{c})$ in the considered
    observing periods; OP3L is the low-luminosity episode.
	The flux is computed
    from the \textsl{ISGRI} data in the 20--40\,keV energy range and expressed in units
    of $10^{-10}\mathrm{erg\,cm^{-2}\,s^{-1}}$.
	The uncertainties are at 90\% c.l.
    }
\begin{tabular}{l c c c c}
\hline
\hline
          &  Flux   & $\Gamma$    & $E_c$ [keV]   & $\chi^2/$d.o.f.\\
          &(20--40\,keV)&        &                &                \\
\hline
 OP1 &  $1.7\pm0.5$ & $0.8\pm1.0$      & $22^{+20}_{-9}$  & 66/66\\
 OP2 &  $2.2\pm0.4$ & $0.9\pm0.6$      & $29^{+16}_{-9}$  & 55/66\\
 OP3 &  $4.9\pm0.3$ & $0.97\pm0.17$    & $27^{+4}_{-3}$   & 104/66\\
 OP3L&  $1.8\pm0.5$ & $2.5\pm0.2$      & $> 150$          & 90/74 \\
 OP4 &  $10.5\pm0.2$& $0.34\pm0.05$    & $20.1\pm0.6$     & 97/66 \\
 \hline
\end{tabular}
\label{tab:best_fit}
\end{center}
\end{table}

The broad-band \textsl{JEM-X, ISGRI}, and \textsl{SPI} (4--110\,keV)
spectra for OP3 and OP4 were fitted first using a cut-off power
law with low-energy absorption (\emph{WABS} model in XSPEC) plus a
Gaussian emission line at $\sim$ 6.5\,keV to model the blended
neutral and ionized iron K-emission. The Gaussian emission line has parameters
given by:
$N_{\text{g}}/\sqrt{2\pi\sigma_\text{g}^2} \exp ( - (E-E_\text{g})^2/4\sigma_\text{g}^2 )$.
Best-fit parameters are shown
in Table~\ref{tab:fit_results} (left panel). The spectra were also
successfully fitted with a Comptonization model (\emph{compTT} in
XSPEC) based on a spherical geometry using the analytical
approximation of \citet{comptt} plus an additive Gaussian emission-line
for the ionized K-emission from Iron and low-energy absorption.
Best-fit parameters for this model are shown in
Table~\ref{tab:fit_results} (right panel).
The compTT model was used with approximation 2
and for red-shift 0. $kT_0$ is the temperature of the seed photons
for the Comptonization, $kT_e$ the temperature of the scattering electron cloud,
while $\tau$ is its optical depth in spherical geometry.
The cut-off
model is slightly better for OP3, the compTT model may be preferred
for OP4.

\begin{figure*}
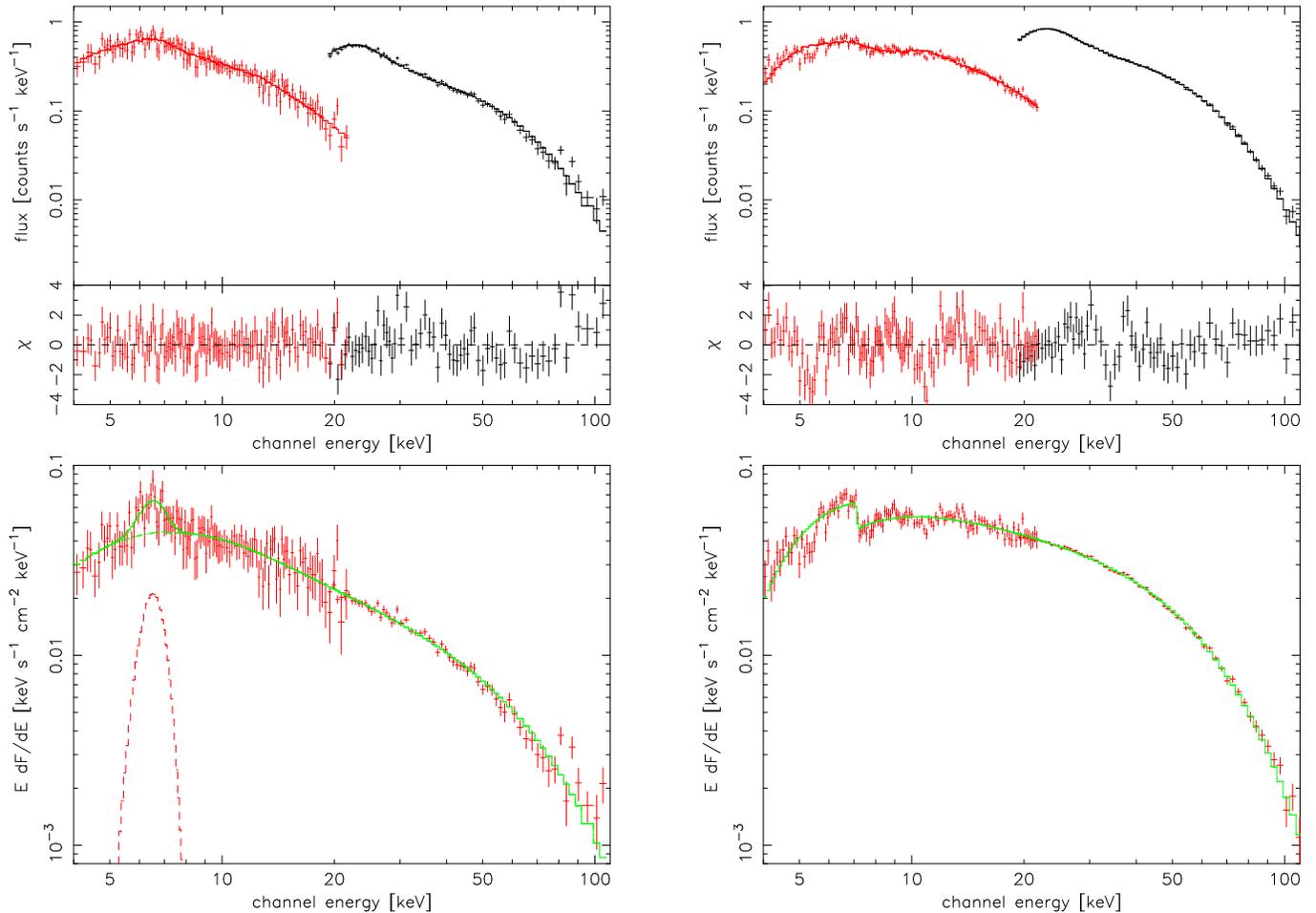

  \begin{center}

   \includegraphics[angle=270,width=8.4cm]{op3_ld.ps} \hfill
   \includegraphics[angle=270,width=8.4cm]{spec_op4_ld.ps}\\
   \includegraphics[angle=270,width=8.4cm]{op3_euf.ps} \hfill
   \includegraphics[angle=270,width=8.4cm]{spec_op4_euf.ps}
    \caption{
    Count-rate spectra (upper panels) and energy-flux unfolded spectra (lower panels)
    for OP3 (left) and OP4 (right) of GX~1+4. The model for the continuum is compTT plus a
    low-energy absorption and Gaussian emission line at $\sim 7$\,keV. The data between 4 and 21\,keV
    are from \textsl{JEM-X}, and the data
    extending from 19 to 100\,keV are from \textsl{ISGRI}.
    }
\label{fig:fit_comptt}
\end{center}
\end{figure*}

Significant differences between the spectra of OP3 and OP4 (at more than 90\% c.l.) are found
in the power law index and cut-off energy (cut-off model) and
in the electron temperature and optical depth (compTT model).
In Fig.~\ref{fig:fit_comptt} we plot the
broad-band count-rate and energy-flux unfolded spectra for OP3 and OP4 using
the compTT model.

The visual inspection of the fit residuals (see
Fig.~\ref{fig:fit_comptt} upper right panel) shows low points around
34\,keV that are reminiscent of a similar structure at a 2.5$\sigma$
level seen at the same energies in \textsl{Beppo-SAX} observations
\citep{rea}. However, the introduction of a multiplicative Gaussian
absorption line $N_{\text{abs}} \exp ( -
(E-E_\text{abs})^2/2\sigma^2 )$ does not improve the fit, since its
significance is much less than $3\sigma$.

The ISGRI spectrum of the low-luminosity episode (OP3L in
Table~\ref{tab:obs}) was also extracted, but \textsl{JEM-X} and
\textsl{SPI} did not detect the source in this state. The spectrum
(left panel of Fig.~\ref{fig:veryweak}) was modelled by a power law
with a slope of $2.5\pm0.2$ giving $\chi^2/$d.o.f.$=90/74$. The
\emph{PEXRAV} model \citep{pexrav} proposed by \citet{rea}, which
includes a reflection component besides the power law, is not
constrained by our data.

\begin{table*}
\begin{center}
\caption{Results for the simultaneous fit of \textsl{JEM-X, SPI} and \textsl{ISGRI} spectra.
    The exposures are reported in Table~\ref{tab:obs} and the uncertainties of
	the best-fit parameters are at 90\% confidence level. The
    physical units of the parameters are reported between squared brackets except for
    the unabsorbed flux, which is computed in the 2--60\,keV energy range and expressed in units of
    $10^{-9}\,\mathrm{erg\,cm^{-2}\,s^{-1}}$.}
\begin{tabular}{|l c c|l c c|}
\hline
\multicolumn{3}{|c|}{Cut-off power law model} & \multicolumn{3}{|c|}{compTT model}\\
parameter & OP3 & OP4 & parameter & OP3 & OP4 \\
\hline
$N_H\; [10^{22}\, \mathrm{cm^{-2}}]$ & $10\pm3$  & $5.0\pm1.5$      & $N_H\; [10^{22}\, \mathrm{cm^{-2}}]$  &    $<10$                 & $25\pm2$        \\
$E_g$ [keV]                          & $6.8_{-0.6}^{+0.3}$ & $6.5\pm0.1$ & $E_g$ [keV]                           & $6.5\pm0.3$         & $6.6\pm0.3$ \\
$\sigma_\text{g}$ [keV]              & $1^{+1}_{-0.5}$      & $<0.4$    & $\sigma_g$ [keV]                      & $0.5\pm0.5$         & $0.2$\,fixed \\
$N_\text{g}\;[10^{-3}\,\mathrm{cts\;cm^{-2}\;s^{-1}}]$  & $7_{-3}^{+6}$ & $2.7\pm0.5$ &  $N_\text{g}\;[10^{-3}\,\mathrm{cts\;cm^{-2}\;s^{-1}}]$  & $4\pm2$     & $1.1\pm0.6$ \\
$E_\text{c}$ [keV]                   & $32\pm4$  & $21.8\pm0.7$      & $kT_0$ [keV] & $1.7\pm0.3$ & $1.6\pm0.2$             \\
$\Gamma$                             & $1.2\pm0.2$ & $0.51\pm0.05$ & $kT_e$ [keV]                    & $15.4\pm1.4$   & $13.1\pm0.2$ \\
              &                 &                                       & $\tau$  & $5.1\pm0.5$  &  $6.80\pm0.15$ \\
Flux 2--60\,keV & $1.77\pm0.07$ & $ 2.32\pm0.03$  & Flux 2--60\,keV & $1.70^{+0.04}_{-0.16}$ & $2.16\pm0.04$ \\
$\chi^2$/d.o.f. & 212/212 & 311/199 & $\chi^2$/d.o.f. & 224/211 & 303/201 \\
\hline
\end{tabular}
\label{tab:fit_results}
\end{center}
\end{table*}

\subsection{Phase resolved spectroscopy}
\label{sec:phase_spectroscopy}
The OSA software (up to the current 5.1 version) does not provide tools for performing
phase resolved spectroscopy for
\textsl{ISGRI} and
\textsl{SPI}. In the case of \textsl{JEM-X} the task is very cumbersome.
Thus we developed dedicated tools (described in
the Appendix~\ref{sec:isgriextract}) to analyse the ISGRI data.

Phase-resolved spectra of OP3 and OP4 were extracted according to
the phase interval selection shown in the upper panel of
Fig.~\ref{fig:phase_spectra} and modelled with a cut-off power law.
In OP1 and OP2 the source was too weak for this kind of analysis. As
far as OP3 is concerned, best-fit parameters do not change with the pulse phase within
the statistical uncertainties. On the
contrary, the best-fit parameters of phase-resolved spectra of OP4
(Table~\ref{tab:phase_resolved}) change significantly with pulse
phase: the power-law slope is not constant at a 99.9\% confidence
level, while the cut-off energy is not constant at a 90\% confidence
level.

\begin{figure}
  \begin{center}
    \resizebox{0.8\hsize}{!}{\includegraphics[angle=270]{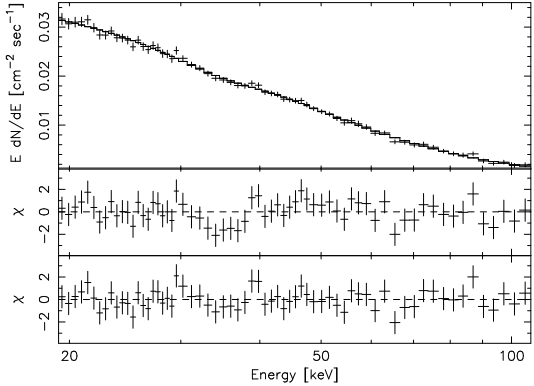} }
    \caption{
    In the upper panel we show the \textsl{ISGRI} energy-flux unfolded spectrum
    extracted in the phase interval 0.6--0.85 The best-fit model has a multiplicative
    Gaussian absorption line at ($34\pm2$)\,keV.
    The other panels show, from top to bottom,
    the residuals without the introduction of the line and the residuals with its introduction.
}
\label{fig:descent}
\end{center}
\end{figure}

Although not formally requested by the $\chi^2$ of the fit,
we also searched for absorption-like features in the phase-resolved spectra.
This was, again, motivated
by the residuals observed in the spectrum of phase 0.6--0.85 (Fig.~\ref{fig:descent}),
the descending part of the pulse,
and by the weak evidence of such a line reported by \citet{rea} and \citet{naik2004}.
Using the Gaussian absorption model (see Sect.~\ref{sec:ph_av_spec}), the line centroid
is found at $34\pm2$\,keV. The line width is not constrained by the fit and is set to 4\,keV,
compatible with the width of several absorption lines observed
at similar energy in other sources \citep{coburn2002}.
The line significance using an F-test is 2.4$\sigma$, but, due to the questionable reliability of this
statistical test in astronomical data analysis \citep{ftest},
we conclude that the evidence for such a line, if any, is weak.
However, systematic effects
can be excluded because such a feature is not present in other pulse phases.

\begin{table*}
\begin{center}
\caption{Fit results for phase-resolved spectra, $\Delta\chi^2$ is referred to the passage
from 72 to 70 d.o.f  with the introduction of a Gaussian absorption line.
The $\sigma$ of the Gaussian absorption line
is fixed at a value of 4\,keV. The line significance is obtained
using the F-test probability value. The uncertainties are the joint confidence limit at 90\% probability.}
\begin{tabular}{l c c c c}
\hline
\hline
phase                 & 0.10--0.35     & 0.35--0.6     & 0.6--0.85                  & 0.85--1.10 \\
\hline
$\Gamma$              & $0.34\pm0.14$ & $0.10\pm0.08$  & $0.4\pm0.1$                & $0.6\pm0.2$    \\
$E_c$ [keV]           & $19\pm1$      & $18.1\pm0.7$   & $22.5\pm1.5$               & $20\pm3$       \\
$E_\text{abs}$ [keV]  &               &                & $34\pm2$                   &                 \\
$N_\text{abs}$        &               &                & $1.11\pm0.05$              &                 \\
$\Delta\chi^2$        &               &                & $69(72)\rightarrow56(70)$  &                \\
line significance     &               &                & $2.4\sigma$                &                 \\
\hline
\end{tabular}
\label{tab:phase_resolved}
\end{center}
\end{table*}

\begin{figure}
  \begin{center}
    \resizebox{0.8\hsize}{!}{\includegraphics[bb=10 66 540 790, angle=0]{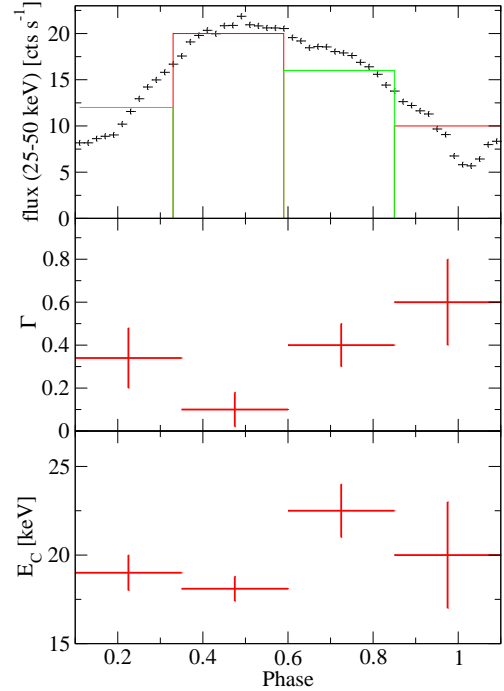} }
    \caption{
    The pulse profile in the 25--50\,keV energy band with the phase selection (upper panel),
    the slope (middle panel), and the cut-off energy (bottom panel) of the phase-resolved spectra of
    ISGRI in OP4.
    The best-fit parameters are obtained for the cut-off power-law model $E^{-\Gamma}\,\exp(E/E_\text{c})$
    without the introduction of the Gaussian absorption line
    in the energy range 17--110\,keV. The uncertainties represent the joint confidence limit
    at 90\% probability ($\Delta\chi^2 = 2.7$).
    }
\label{fig:phase_spectra}
\end{center}
\end{figure}

\section{Discussion}
\label{sec:discussion}
The main results of our analysis of the \textsl{INTEGRAL} observations of GX~1+4 are:

\begin{enumerate}

\item The measurement of the longest spin period
in the observational history of GX 1+4: $P_\mathrm{s}\sim$141\,s;

\item The measurement of the secular spin-down trend $\dot P~=~(6.6 \pm 0.8)\times10^{-8}$\,s/s,
in agreement with previous findings;

\item The detection of a variation from
spin-down to spin-up associated with the high luminosity state;

\item Weak evidence in the phase-dependent spectra of an absorption feature with fundamental energy at
$33~-~35$\,keV that could be due to cyclotron resonant scattering;

\item The observation of a low-luminosity episode
where the source is found pulsating and the spectrum does not show evidence of any
reflection component.
\end{enumerate}
\noindent
In our Integral analysis we have also found, in line with
previous observations from other satellites, evidence of energy
dependent pulse profiles, spectral variations between luminosity
states, and phase dependent spectra.

We confirm that the source is steadily spinning-down and maintaining a
much lower luminosity than in the 1970s. The values of the measured
spin-period derivative fall in the same range obtained on the
basis of \textsl{BATSE} data. However, we observed a spin-up of the source in the high 
luminosity state: we exclude the possibility that the spin
derivative inversion is due to a binary-system-induced Doppler shift,
as already demonstrated by \citet{pereira1999}; it is more likely a
signature of a torque-reversal episode, which is rather common in GX~1+4
\citep{chakba1997}. In our observation the torque-reversal episode
happened at a luminosity level compatible to the ones that
\citet{chakba1997} reported as occurring at MJD~49\,300, 49\,700, and
50\,300. Unfortunately, on the basis of our \textsl{INTEGRAL}
observation, it is not possible to narrow down the period at which
the torque reversal occurred. It also needs to be stressed that a
torque reversal does not imply such a luminosity value and
vice-versa as can be seen from Fig.~2 of \citet{chakba1997}, where
one sees spin-up episodes at lower luminosity and strong
luminosity increases without a corresponding spin-up.

Considerations of the nature of the emission mechanism can be made by
discussing the pulse-profile evolution with energy, together with the
phase resolved spectroscopy. 
When the source is brighter ($L_{20-40\,\mathrm{keV}}= 10.5 \times
10^{-10}\mathrm{erg\,cm^{-2}\,s^{-1}}$, OP4),
the pulse profile can be heuristically described by two Gaussian sub-peaks centred at
phases $\sim0.4$, $\sim0.7$: the relative intensity
between the first and the second peaks increases with energy (Fig.~\ref{fig:op4_pulses}).
At $\sim20$\,keV the pulse appears more peaked at an early phase while
at energy just below 80\,keV the comparison with Fig.~\ref{fig:pulses} shows
that the two sub-peaks have equal intensity and the pulse shape becomes similar to the lower
luminosity states observed in OP2 and OP3.

As the pulse shape changes with energy, the spectral properties evolve
with phase: in the ascending and top part of the
pulse the spectrum is well-fitted with a power law of slope $\sim
0.2$ and cut-off energy $\sim 19$\,keV, while in the descending part
the spectrum softens. This suggests a larger optical depth for scattering
in the first sub-peak with respect to the second sub-peak, which turns out to be more
similar to the phase-averaged spectrum of OP3.
Such a phase-dependent behaviour is different from the one observed
in previous phase-resolved analyses
\citep{galloway2000,galloway2001,naik2004}, which are focused on an
absorption enhancement in correspondence with the pulse dip (at phase
zero in our pulses) and do not find significant changes in the
continuum within the remaining part of the pulse.

The broad-band \textsl{JEM-X/ISGRI/SPI} spectra are modelled either
with an absorbed cut-off power law or with a Comptonization model.
In OP1 and OP2, the fit parameters are not well-constrained. In OP4
(when the source is brighter), we find a harder emission up to
$\sim40$\,keV, followed by a faster decay, in comparison to OP3. In
contrast to our results, \citet{cui2004} find that when the
source is bright ($\sim 10^{-9}\,\mathrm{erg\,cm^{-2}\,s^{-1}}$ in
the 2--60\,keV energy band), the spectral index is very stable
between 1 and 1.3 with $e$-folding energy at 28--40\,keV. They also
find that the fit parameters vary erratically with no apparent
correlation to the flux on a time scale of a few days when the source is
dimmer (spectral index between 0.2 and 2 and $e$-folding energy
between 11 and 45\,keV). These considerations imply that the spectral
properties of GX~1+4 are not determined uniquely by its luminosity
state, but also depend on some still unknown parameter.

The picture is slightly clearer when we use a thermal Comptonization
model: the value of the seed component temperature
($kT_0\simeq1.6$\,keV) is always similar to the one found with
\textsl{Beppo-SAX} and \textsl{RXTE}, while the optical depth and
plasma temperature are about the same but cannot be
straightforwardly compared, confirming the variability of the accretion
mechanism within a stable system. Limiting ourselves to our data,
where the source is brighter, the optical depth is higher with
respect to the lower luminosity state, while the temperature of the
Comptonization plasma is lower. This spectral evolution with
luminosity is consistent with the thermal Comptonization being the
dominant emission process \citep[see e.g. the conclusions in
][]{becker2005}: using this model the larger accretion rate can
explain the enhancements of both luminosity and optical thickness in
the accreting plasma.

We observe weak evidence (at 2.4$\sigma$ level) of an absorption
feature at $34\pm2$\,keV in the descending part of the pulse
profile, at phase [0.6--0.85] (see Fig.~\ref{fig:descent}).
Interpreting the absorption as due to the cyclotron resonant scattering
of electrons, the magnetic field in the emitting region would be
$(2.9\pm0.2) \times 10^{12} (1+z)$\,G where $z$ is the gravitational
red shift of the emitting region.
Its presence or enhancement just in the descending part of the pulse
is typical of other systems such as 4U~0115+63
\citep{santangelo1999,heindl1999} or Cen~X-3 \citep{santangelo1998}.
The inferred magnetic field would be one order of magnitude less
than expected from the standard accretion torque theory, but
consistent with the one inferred from the model of emission due to a
retrograde disk \citep{chakba1997}. Although the calibration
uncertainties of \textsl{ISGRI} suggest that this weak
evidence is an indication rather than a measurement, we note that a
similar indication comes from PDS on board of \textsl{Beppo-SAX}. On
the other hand,  it is not surprising that \textsl{RXTE}
observations did not show spectral structures, although observations
were carried out many times and in a variety of states
\citep{giles2000,cui2004}. The single observations were shorter than
those by \textsl{INTEGRAL}, and it is not possible to add many
observations together because of the spectral variability of the
source.

The low-luminosity episode observed with \textsl{INTEGRAL} is similar to
the ones seen by \textsl{Beppo-SAX} \citep{rea,naik2004} and \textsl{RXTE}
\citep{galloway2000low}.
The \textsl{Beppo-SAX} observation of such a low-luminosity episode
showed a pulsed signal and a spectrum with an exponential
roll-over.
\citet{cui2004} found with \textsl{RXTE} that, when the pulsation disappears or is very weak,
the spectrum can be modelled by a simple power-law with index $\sim$1.6.
They also find episodes of
suppressed flux that show a pulsation and a detectable roll-over in the spectrum.
During the low-luminosity episode observed by \textsl{INTEGRAL},
the source is found to pulsate,
but a simple power law with a slope of $2.5\pm0.2$
models the spectrum of the source up to 150\,keV without the need of an exponential cut-off.
This seems in contrast with \citet{cui2004}.
On the other hand, we were not able to observe any signature of emission
from a disk as found by \citet{rea} or an enhancement of the absorption as found by \citet{naik2004};
this is due to the relatively high (17\,keV) lower
threshold of our spectrum. In conclusion, with the analysed \textsl{INTEGRAL} observations we cannot
settle the puzzle regarding the origin of these low-luminosity
states.

\section{Conclusions}
\label{sec:conclusions}
In this paper we analysed the \textsl{INTEGRAL}
data of the low-mass X-ray pulsar GX~1+4. The source was observed
several times and found in
different luminosity states with significant variations in its
spectral and timing characteristics.

The overall spin-down trend is confirmed by the highest spin
period ever observed for this source and by the negative local
spin derivative we measured during most of the observation. However, in
the high-luminosity state GX~1+4 showed a local
spin-up. We estimate that the torque reversal happened at a
source luminosity similar to analogous episodes described in
\citet{pereira1999}. We notice that, during the spin up phase,
the pulse profile is more peaked than in the intermediate
luminosity state. 

When the source was brighter, the spectrum showed a sharper exponential cut-off
after becoming harder between 10 and 50\,keV. This is consistent with the
thermal Comptonization being the dominant emission process.

We also observed weak evidence of an absorption feature in the phase-resolved
spectrum of GX~1+4 at $33$--$35$\,keV in the phase interval [0.6--0.85].
\citet{rea} report a similar evidence in the phase-averaged spectrum obtained with \textsl{Beppo-SAX}.
If this feature is due to cyclotron resonant scattering,
the pulsar magnetic field would be $\sim 3\times 10^{12}$\,G.

\section*{Acknowledgments}
The observational data used in this communication were collected by
INTEGRAL, an ESA science mission for X-ray and Gamma-ray astronomy.
The work was supported by the Italian Space Agency (ASI) under
contract no.~I/R/046/04 and ASI-INAF I/023/05/0, and by the German Space Agency (DLR) under
contract nos.~50~OG~9601 and 50~OG~0501. We thank Teresa Mineo and
Nikolai von Krusenstiern for many useful comments on the manuscript.
We also thank the anonymous referee for valuable comments and
suggestions.
\nocite{segreto2006}
\bibliographystyle{aa}
\bibliography{pulsars}

\appendix
\section{ISGRI energy calibration issues}
\label{sec:calibration}
The \textsl{ISGRI} detection layer consists of a 128$\times$128 array of
independent CdTe detector pixels \citep{isgri}.
This kind of detectors is affected by the ``Charge
Loss Effect'': since in CdTe the hole lifetime is much shorter than its transit time,
the induced current can be significantly reduced depending on the position of the interaction on the detector.
To reconstruct the energy of the incoming photons,
it is necessary to perform an energy
correction of the pulse amplitude value as a function of the pulse rise time; for this purpose
OSA software makes use of
multiplicative coefficients stored in a ``Look-Up'' Table (LUT2).

\begin{figure}
  \begin{center}
    \resizebox{\hsize}{!}{\includegraphics{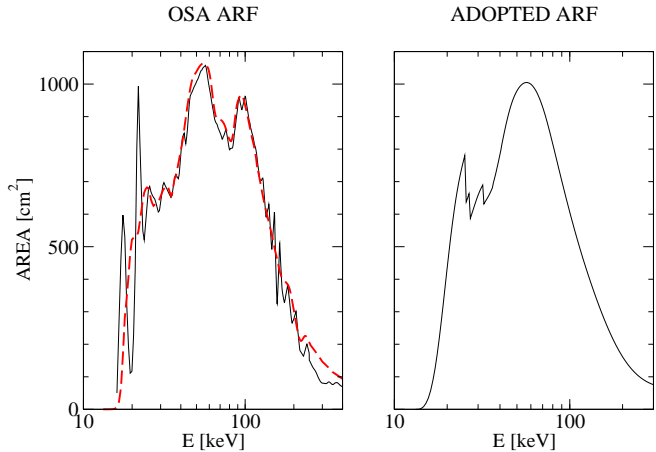}}
    \caption{
    Comparison between the effective area as distributed with OSA (left panel)
      and the new effective area obtained after
      performing the energy correction with the new LUT2 (right panel).
      In the left panel the dashed line is the effective area
      distributed with OSA~4.2, and the solid line the effective area distributed with OSA~5.1.
      The OSA effective areas present many
      bumps and wiggles, which are introduced ad hoc to compensate for the OSA LUT2 energy
      correction. Features in the 30--40\,keV region are due to CdTe absorption edges.
     }
\label{fig:arf}
\end{center}
\end{figure}

The LUT2 distributed with the OSA software
is not optimized, especially
in an energy region near 80\,keV where a fraction of the
photons is overcorrected.
Ad hoc absorption-like features at $\simeq80$\,keV have been introduced in the
OSA~4.2 and OSA~5.1 \textsl{ISGRI} Effective Area (respectively dashed and solid lines in  Fig.~\ref{fig:arf})
modelled on the basis of the observed Crab spectrum assumed to be a power law
with index 2.2. This choice of the index does not agree
with the widely accepted value of 2.1 \citep{crab1974}.
\textsl{ISGRI} spectra are
thus systematically softer than other instruments.
Moreover, the introduction of these ad hoc corrections in the effective area
could produce the appearance of fake features in the
spectra of sources that are different from the Crab.

To improve the calibration, a new LUT2
based on on-ground and in-flight calibration data has been generated.
Using this new LUT2 a
better charge loss correction is obtained.\footnote{For the technical details we refer to
Alberto Segreto's talk at the Internal \textsl{INTEGRAL}
workshop available at\\
\texttt{http://www.rssd.esa.int/Integral/workshops/Jan2005/} and to the new calibration files available at\\
\texttt{http://www.ifc.inaf.it/\~ ferrigno/integral/}.}

We also generated new spectral-response matrices taking into account the charge loss
effect and its correction by the new LUT2.
The effective area, shown as the
dash-dotted line in Fig~\ref{fig:arf}, does not exhibit any 80\,keV absorption-like
feature. This constitutes a confirmation
of the better energy correction performed
with the new LUT2: there is, in fact, no physical reason for this feature.

\begin{figure}
  \begin{center}
\resizebox{\hsize}{!}{\includegraphics[angle=270]{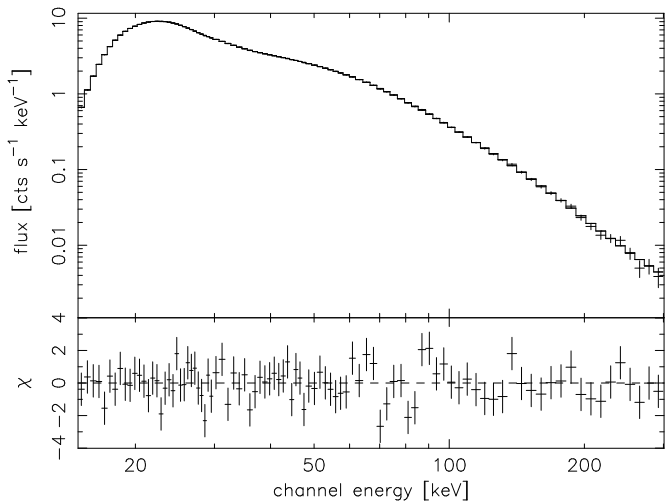} }
    \caption{Crab spectrum from a selection of science windows in the revolutions
    102 and 103 in staring observations, the exposure is about 47\,ks.
    The fit is obtained with a power law (slope 2.1) without adding systematic errors.
}
\label{fig:crab}
\end{center}
\end{figure}

Also the spectral parameters we obtained with our response matrices
are closer to the literature values (e.g. an index 2.1 for the Crab), while
the $\chi^2$ values of the fits are systematically better than the ones obtained
using the OSA~4.2 and 5.1 calibration files.
In Fig.~\ref{fig:crab} the fit to the Crab spectrum can be appreciated, but the reader
can read a more detailed discussion in the appendix of
\citet{mineo2006}.

If one assumes that the fractional spread of the Crab residuals of Fig.~\ref{fig:crab} is
an empirical indicator of the calibration systematic errors,
the calibration systematics can be neglected for our observation in OP4. In fact, the Crab reference spectrum we used
and the source spectrum have about the same number of total counts in the 20--40\,keV energy range
but the S/N of the Crab is three times higher.

\section{Timing analysis of the ISGRI data.}
\label{sec:isgriextract}

The standard OSA pipeline does not allow to perform timing analysis of
pulsars. We then developed an alternative
software\footnote{The software can be downloaded at the URL\\
\texttt{http://www.ifc.inaf.it/\~ferrigno/INTEGRALsoftware.html}.}
to extract simultaneously spectra, light curves and phase resolved
spectra from the event lists; a dedicated paper which describes
the method in detail (Segreto et al., in prep.) is in preparation,
here we give a brief overview.

When illuminated by an X-ray source, a coded mask casts a
shadow on the detector plane (shadowgram) that, correlated with
the known mask pattern, allows to obtain the position and
intensity of the sources within the instrument field of view
\citep{goldwurm2003}.

To obtain spectra and light curves of these sources,
a shadowgram deconvolution process can be repeated
several times, each time after selecting events in the energy and time
intervals of interest.
However, when narrow energy or time bins are considered, a large amount of
CPU time is required.

A much faster way to obtain light curves or
spectra from coded mask telescopes when the source positions have
already been determined (because known a priori or predetermined
by a shadowgram deconvolution), is via the use of the
illumination pattern that
the mask
casts on the detection plane. Given the pixellated nature of the
detector plane, some of the detector pixels will then be fully
illuminated by a source while other pixels will be partially
illuminated or not illuminated at all. For each pixel the amount
of illumination from each source, normalized to the maximum
illumination value is called the pixel Photon Illumination Fraction (PIF).

Using this PIF information, spectra, light curves, or other
scientific products can be extracted by means of a minimum amount of algebraic
operations, without the need of repeating a shadowgram
deconvolution process over and over again. This method, which is used e.g. by
the standard pipeline of BAT \citep{BAT} on board Swift
\citep{SWIFT} and by the \textsl{JEM-X} software, does not account for the presence of
multiple sources in the same field of view.
Due to a non ideal imaging point-spread function this may produce
cross contamination between the signals from
different sources.
To avoid this problem, we then developed a decontamination algorithm,
based again on simple algebraic operations \citep{segreto2006}.

To validate our extraction method we
compared the background-subtracted spectra of several
sources with the one obtained using the OSA standard
pipeline and verified that the two methods give equivalent results
not only in terms of spectral shape but also in
their S/N.

The PIF-based extraction method we developed is not only much faster, but it also allows to easily perform
fast timing analysis and phase-resolved spectroscopy, tasks not supported by the
standard OSA software.

\begin{figure}
  \begin{center}
    \resizebox{\hsize}{!}{\includegraphics[angle=270]{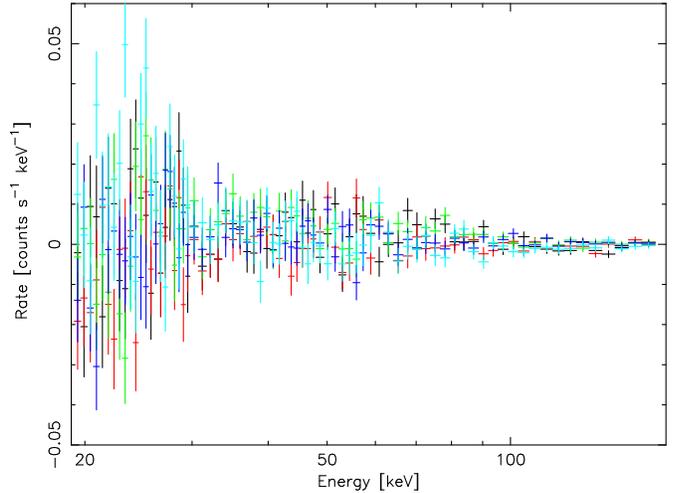} }
    \caption{
    Spectra extracted in five positions at two degrees from GX~1+4 for the
    OP4 data set.
    }
\label{fig:back}
\end{center}
\end{figure}

To estimate the reliability of the background rejection we extracted
spectra in several sky directions
chosen around the main source to ensure the same observational conditions but
far enough to avoid problems due to the $12'$ point-spread function \citep{gros2003}.
In Fig.~\ref{fig:back} we plot an example for five pointings within OP4.
As shown, the spectra
are all consistent with zero without points with S/N exceeding $\sim3$ in each spectrum.
We verified that the S/N
distribution of the spectra in the 19--110\,keV range
is described well by a Gaussian function with $\sigma\sim1$.

\end{document}